\documentclass{webofc}

\usepackage[varg]{txfonts}   
\usepackage{hyperref}
\usepackage{url}
\hypersetup{colorlinks=true,citecolor=blue,urlcolor=blue,linkcolor=blue}
\begin{document}
\title{Recent progress in global optimizations of covariant energy density functionals}
%
%

\author{\firstname{A. V. } \lastname{Afanasjev}\inst{1}\fnsep\thanks{\email{Anatoli.Afanasjev@gmail.com
       }} \and
        \firstname{B.} \lastname{Osei}\inst{1}\fnsep\thanks{\email{bo344@msstate.edu}} \and
        \firstname{A.} \lastname{Dalbah}\inst{1}\fnsep\thanks{\email{alidalbah17@gmail.com}}
}

\institute{Department of Physics and Astronomy, Mississippi State University, Mississippi 39762, USA}

\abstract{The recent progress on global optimizations of covariant energy density functionals (CEDFs)
and global calculations of binding energies within the covariant density functional theory (CDFT) has 
been analyzed and reviewed.  Recently developed anchor-based optimization  approach of
Ref.\ \cite{TA.23} allows global optimizations of CEDFs at a reasonable numerical cost. 
Moreover, it permits such optimizations  in a very large fermionic basis with a proper extrapolation
to an infinite one. This allows to accurately estimate global calculation errors due to use of truncated 
fermionic basis and neglect of some contributions to binding energies (such as total electron binding energy).}
%
\maketitle
%
\section{Introduction}
\label{intro}

   Nuclear binding energies (or masses) is one of the most fundamental properties of 
atomic nuclei the accurate knowledge of which is important for nuclear structure, neutrino 
physics and nuclear astrophysics.  In experiment, they are measured with an accuracy better 
than 1 keV.  Theoretical calculations do not reach such an accuracy but there is on-going and 
continuous effort of different nuclear theory groups on an improvement of global description of
nuclear masses.  They are guided not only by the need for such predictions in nuclear 
astrophysics but also by an understanding that such observables, which are accurately
and globally measured, provide important constraints on the properties of nuclear energy
density functionals.

   The success in theoretical calculations of binding energies has been quite impressive 
in non-relativistic microscopic+macroscopic (mic+mac) approach and  density functional 
theories (DFTs)  with an accuracy of the global reproduction of experimental masses reaching 
$\approx 0.5$ MeV. One should say that it is achieved by going beyond the pure 
DFT framework via  the introduction of phenomenological vibrational+rotational corrections (VRC)  
and  Wigner energy. Without such corrections (i.e. at the mean field or at the DFT level) the accuracy
of the reproduction of experimental data is seldom better than 1.5 MeV but typically it is lower that 
this value (see, for example, Ref.\ \cite{UNEDF2}).  A recent study within generator coordinate method
(GCM) revealed some differences between the GCM results obtained for VRC and phenomenological
ones based on  cranking approximation (see Ref.\ \cite{LZY.25}). It is limited to the Cr isotopes: thus, the 
global  accuracy of phenomenological treatment of VRC is still not well established. There are also 
conflicting approaches to the definition of the Wigner energy. For example, this energy is proportional 
to a relative neutron excess $|N-Z|/A$ in the mic+mac  calculations of Ref.\ \cite{MNMS.95}. In contrast, 
it is localized in the vicinity of the $N=Z$ line in Skyrme DFT calculations by Brussels group (see Ref.\ 
\cite{GSHPT.02}).

   A similar progress in the study of binding energies within the CDFT framework  has been 
substantially slower due to the  following factors. First, for a given fermionic basis the numerical 
calculations of matrix elements and the diagonalization of the matrices  requires substantially 
more time (by at least one order of magnitude)  than in  the case of nonrelativistic DFTs since 
Dirac spinors contain large and small components of opposite parity (see Ref.\ \cite{TOAPT.24}). 
Second, binding energies show very slow convergence as a function of the size of fermionic basis 
so that accurate calculations with an evaluation of infinite basis corrections to binding energies
are numerically very expensive (see Refs.\ \cite{TOAPT.24,NL5Z-DDMEZ-PCZ}). It is only
with the introduction of anchor-based optimization approach in Ref.\ \cite{TA.23} that
such global optimizations of CEDFs become feasible (see Ref.\ \cite{NL5Z-DDMEZ-PCZ}).
Recent global optimizations of  harmonic oscillator basis in the CDFT carried out in
Ref.\ \cite{OAD.25} offer the path to a further reduction of numerical expenses.

  The present paper analyses and reviews global optimizations of CEDFs and
global mass calculations carried out in the CDFT. Using the 
functionals fitted in Ref.\ \cite{NL5Z-DDMEZ-PCZ} as a benchmark, it discusses
global calculation errors. Finally, the bottlenecks for the improvement of the
description of binding energies are identified and traced back to specific regions of
nuclear chart. They are due to the deficiencies in the description of single-particle
properties.

\section{The status of global calculations}
     
    Global mass calculations in the CDFT have been carried for almost three 
decades and  Table \ref{table-global-mass} provides a summary of such calculations 
for which $\Delta B_{rms} \leq 3.0$ MeV. Here $\Delta B_{rms}$ is the rms difference 
between calculated  and  experimental  binding energies. Column 3 displays whether 
corresponding functional has been fitted globally (Yes) or not (No) and whether global
fits were done at the  mean field (MF) or beyond mean field (BMF) levels. Note that
only in Ref.\ \cite{AGC.16} the latter was employed but the BMF effects were treated 
at the phenomenological level (PHC) via rotational and vibrational corrections.  Column 6 
indicates whether the BMF effects have been into account in the global calculations either 
at the PHC or 5-dimensional Collective Hamiltonian (5DCH) levels. The accuracy of the 
description of binding energies provided in the last column of  Table \ref{table-global-mass} 
is discussed in Sec.\ \ref{global-error}.

  In the CDFT calculations all fermionic and bosonic states belonging to  the shells up
to $N_F$ and $N_B$ are taken into account in the diagonalization of the Dirac 
equation and the matrix inversion of the Klein-Gordon equation, respectively (see
Sec. III in Ref.\ \cite{TOAPT.24} for details).  These values are shown in the column 4 of 
Table \ref{table-global-mass}. Note that Woods-Saxon (WS) basis is used in  Ref.\ 
\cite{RHB-e-e-continuum.22} and it is truncated in a different way as compared with 
harmonic oscillator (HO) basis. The truncation of the HO and WS bases leads to numerical 
errors in the calculation of nuclear binding energies. It is only in Ref.\ \cite{NL5Z-DDMEZ-PCZ} 
that the approach effectively approximating to infinite basis is used in the calculations.

    Column 5  in Table \ref{table-global-mass} indicates whether total electron binding 
energies (TEBE) have been taken into account in the conversion of experimental atomic binding energies 
into nuclear ones. Binding energies provided in the Atomic Mass Evaluations (AME) are atomic ones
i.e. they are the sums of nuclear binding energies and total electron binding energies \cite{AME2020-second}.
The absolute values of TEBE gradually increase from almost zero for $Z=2$ to approximately 1.6 MeV in
superheavy nuclei with $Z\approx 120$ (see Fig. 1 in Ref.\ \cite{NL5Z-DDMEZ-PCZ} and the last column 
in Table III in Ref.\ \cite{DFA.24}). TEBE is neglected in absolute majority of fitting protocols 
of energy density functionals (EDFs) and global mass calculations with these EDFs
(see Sec.\ III in \cite{NL5Z-DDMEZ-PCZ}). This represents another source of global errors.

\begin{table}[htb]
\centering
\caption{The summary of existing global mass calculations in the CDFT. Note that the 
$\Delta B_{rms}$ values shown in plain in the last column are defined from the comparison of 
experimental atomic binding energies and calculated nuclear ones (i.e. total electron binding
energies are neglected). In contrast, bold style is used for the $\Delta B_{rms}$ values defined
from the comparison of experimental and calculated nuclear binding energies.
 See text for further details.
}
\begin{tabular}{c|c|c|c|c|c|c}
\hline 
  functional                     &     refe-         & global          &     HO    & TEBE          & BMF-   & $\Delta B_{rms} \pm \delta B_{rms}^{negl} $  \\ 
                                      &    rence         &  fit                &       basis        &           & global         &  [MeV]  \\ \hline
       1                             &       2                     &        3               &          4                                  &        5           &      6           &         7                                  \\  \hline                     
NL3*                     &    \cite{AARR.14}  &        No              &         $N_F=N_B=20$    &       No           &      No           &     3.00 \\
DD-ME2               &    \cite{AARR.14}  &        No              &         $N_F=N_B=20$    &       No           &      No           &     2.45 \\
DD-ME$\delta$    &    \cite{AARR.14}  &        No              &         $N_F=N_B=20$    &       No           &      No           &     2.40 \\
DD-PC1                &    \cite{AARR.14}  &        No             &         $N_F=N_B=20$    &       No          &      No            &     2.15 \\
DD-MEB1                   &  \cite{AGC.16}         &       Yes (BMF)             &         $N_F=16$       &       No                     &       PHC             &     1.167  \\
DD-MEb2                   &  \cite{AGC.16}         &        Yes (BMF)             &         $N_F=16$       &       No           &     PHC               &    1.153 \\
PC-PK1    &    \cite{ZNLYM.14}    &        No             &          $N_F=12, 14,16$   &       No           &      No                  & 2.58 \\
PC-PK1    &    \cite{LLLYM.15}     &        No             &        $N_F=12, 14,16$   &       No           &      5DCH                &  1.14 \\
PC-PK1     &  \cite{YWZL.21}   &        No             &            $N_F=12, 14,16$  &       No           &      No                       &   2.64 \\
PC-PK1     &  \cite{YWZL.21}    &        No             &     $N_F=12, 14,16$             &     No            &     5DCH                     &     1.31 \\
PC-PK1     &    \cite{RHB-e-e-continuum.22}  &        No           &        WS basis                 &       No                   &       No             &    2.744 \\
PC-PK1   &  \cite{RHB-e-e-continuum.22}    &        No             &        WS basis        &       No           &     PHC                       &    1.518 \\
    NL5(Y)    &  \cite{TA.23,NL5Z-DDMEZ-PCZ}       &        Yes (MF)            &         $N_F=N_B=20$      &       No           &     No                      &     $2.407  \pm {\bf 0.84}$          \\
    PC-Y    &  \cite{TA.23,NL5Z-DDMEZ-PCZ}        &        Yes (MF)             &           $N_F=N_B=20$       &       No           &     No                       &     $1.951  \pm (>1.0)$            \\
   DD-MEY    &  \cite{TA.23,NL5Z-DDMEZ-PCZ}   &       Yes (MF)            &            $N_F=N_B=20$  &       No           &     No                     &     $1.802  \pm {\bf 0.77}$            \\
  NL5(Z)    &  \cite{NL5Z-DDMEZ-PCZ}          &        Yes (MF)           &        infinite       &       Yes           &     No          &       ${\bf 2.357 } \pm {\bf 0.023}$            \\
 PC-Z    &  \cite{NL5Z-DDMEZ-PCZ}            &        Yes (MF)            &         infinite        &       Yes           &     No           &   $ {\bf 1.901}  \pm {\bf 0.05}$            \\
DD-MEZ    &  \cite{NL5Z-DDMEZ-PCZ}    &        Yes (MF)            &        infinite        &       Yes           &     No          &         $ {\bf 1.557}  \pm {\bf 0.028}$            \\
\hline 
\end{tabular} 
\label{table-global-mass}
\end{table} 

\section{New generation of covariant energy density functionals}
\label{new-generation}
 
   One can see in Table \ref{table-global-mass} that prior to our calculations the global fit of covariant 
energy density functionals (CEDFs) has been carried on only in Ref.\ \cite{AGC.16}. However, because 
of the use of traditional fitting protocol (TFP) such fit was possible only in the $N_F=16$ basis. It is only with 
advent of anchor-based optimization approach (ABOA) in Ref.\ \cite{TA.23} it became possible to carry out 
CEDF optimizations with approximate procedure of extrapolation of nuclear binding energies to infinite 
fermionic basis (see Refs.\ \cite{TOAPT.24,NL5Z-DDMEZ-PCZ}). TFP requires several thousands of the rounds of 
global calculations in axially deformed relativistic Hartree-Bogoluibov (RHB) code. This can be achieved only 
for substantially truncated basis. In contrast, because of fast convergence the fitting protocol built on ABOA 
requires only limited (of the order of 10) number of the rounds of global calculations (see Supplemental Material 
to Refs.\ \cite{TA.23,NL5Z-DDMEZ-PCZ} and Sec. IV in Ref.\ \cite{TOAPT.24}). This allows for the first time in the 
CDFT framework to  define with high accuracy (see Sec.\ \ref{global-error}) the binding energies corresponding 
to infinite fermionic basis using  iterative procedure discussed in Supplemental Material of Ref.\ \cite{NL5Z-DDMEZ-PCZ}.

\begin{figure*}[htb]
\centering
\includegraphics[width=9.0cm]{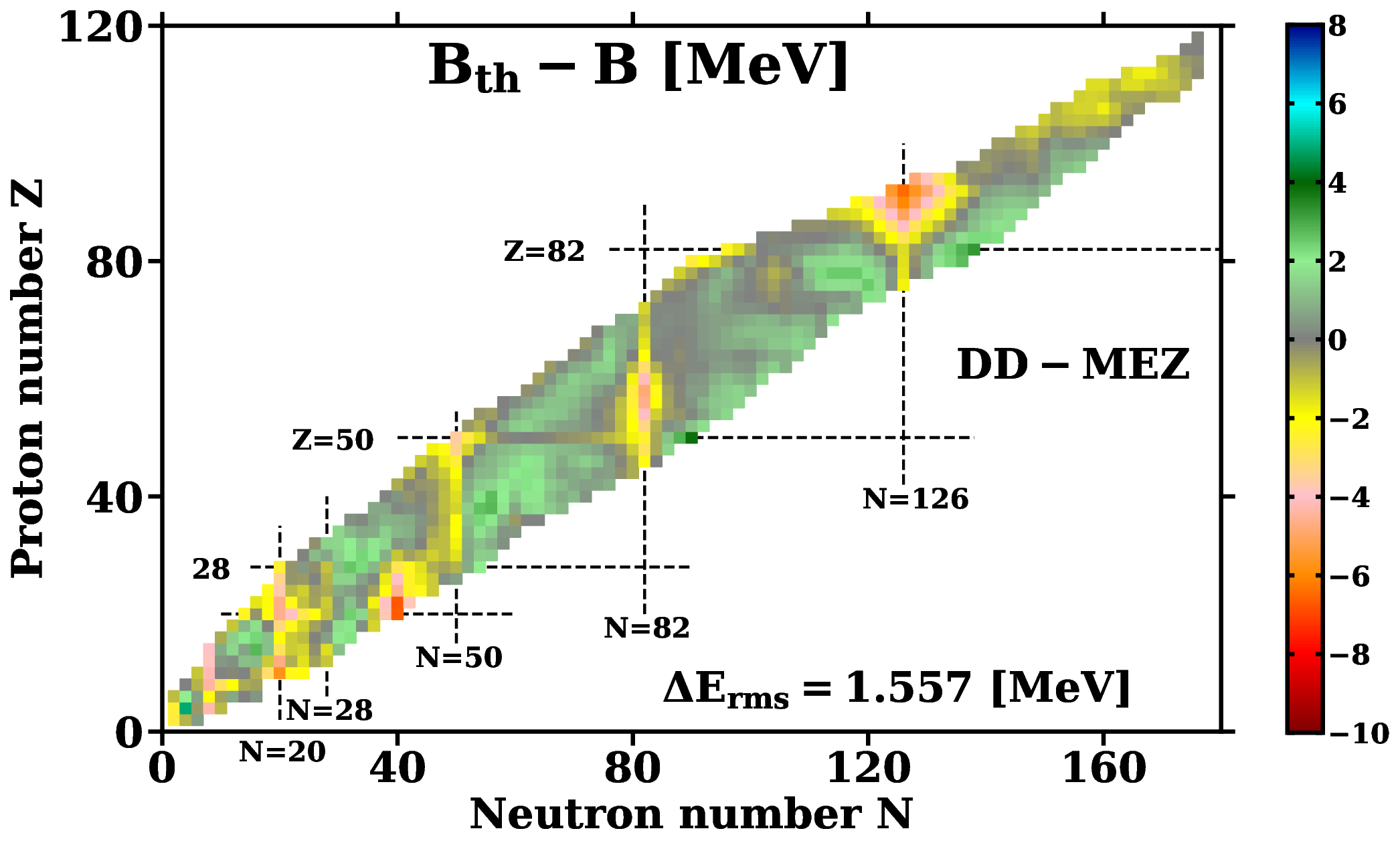}
  \caption{The differences $B_{th}-B$ between calculated ($B_{th}$) and experimental ($B$) nuclear binding energies 
for the DD-MEZ functional. Positive (negative) value of this difference means that calculated nucleus is less (more) bound
than experimental one.
\label{fig-ddmez}
      }
\end{figure*}

   The accuracy of  approximate procedure for extrapolation of nuclear binding energies to infinite 
fermionic basis has been benchmarked in Ref.\ \cite{NL5Z-DDMEZ-PCZ} using significantly improved spherical and 
axially deformed RHB codes which allow the calculations with $N_F=120$ and $N_F=40$, respectively.  It is 
around 10 keV for experimentally known nuclei. Similar analysis for bosonic sector of the CDFT shows that 
the $N_B=40$ basis reproduces the $N_B=120$ results [which corresponds to infinite bosonic basis] with accuracy 
better than 10 keV across whole experimentally known nuclear landscape. The $N_B=20$ bosonic basis was 
used in absolute majority of existing  CDFT calculations during more than thirty years. However, the difference 
between nuclear binding energies calculated with $N_B=20$ and $N_B=40$ reaches 300 and 800 keV in 
superheavy nuclei for the density dependent meson exchange (DDME) and non-linear meson exchange (NLME) 
functionals, respectively.  Thus, we strongly recommend the use of the $N_B=40$ basis in future CDFT calculations 
since it eliminates numerical uncertainties in bosonic sector. Note that contrary to fermionic basis the numerical cost 
of the transition from $N_B=20$ to $N_B=40$ in the bosonic basis is very small.

   Using this numerically accurate calculation scheme and taking into account total electron binding
energies in the conversion of atomic binding energies presented in AME to nuclear ones, the $Z$ class of the functionals 
which includes DD-MEZ, NL5(Z) and PC-Z ones has been developed in Ref.\ \cite{NL5Z-DDMEZ-PCZ}. These two 
features are taken into account in the global mass calculations for the first time in the CDFT framework. Note also 
that in all other functionals presented in Table \ref{table-global-mass} TEBEs are not taken 
into account. This means that  for these functionals the $\Delta B_{rms}$ values presented in the last column of this table are
based on the comparison between experimental atomic binding energies presented in AME and calculated  
nuclear binding energies\footnote{Such contradiction does not exist only in the Z class of the functionals.}. 
The impact of TEBE is only partially taken into account via the redefinition of the parameters in the fitting protocols 
of other CEDFs.

   The DD-MEZ functional provides the best agreement between theory and experiment among the 
functionals fitted at the MF level  and comes very close to those defined at the BMF level (see Table 
\ref{table-global-mass}). Its global performance is illustrated in Fig.\ \ref{fig-ddmez}.

\section{Global calculation errors}
\label{global-error}

  The accuracy of global description of binding energies is typically 
defined  by $\Delta B_{rms}$  (see Table \ref{table-global-mass}). However, such 
an approach ignores the fact that there are global calculation errors 
$\delta B_{rms}^{negl}$ which emerge  from the truncation of 
fermionic and bosonic bases and the neglect of  other contributions to binding energies 
(such as total electron binding energy in the 
conversion from experimental atomic to nuclear binding energies). Thus, strictly 
speaking, the accuracy of the description of experimental nuclear binding energies 
in model calculations has to be characterized by
\begin{eqnarray}
\Delta B_{rms}^{tot} = \Delta B_{rms} \pm \delta B_{rms}^{negl}. 
\end{eqnarray}    
The definition of $\delta B_{rms}^{negl}$ requires the calculations in very large bases 
accurately approximating infinite ones (or reliable evaluation of infinite basis correction 
to binding energies) and so far this has been done only for the Z class of the functionals 
(last three lines of Table \ref{table-global-mass}).  In contrast, this error was not defined 
for other functionals presented in this table since the calculations with them were carried 
out in finite fermionic and bosonic bases and total electron binding energies were 
neglected.

\begin{figure*}[htb]
\centering
\includegraphics[width=8.0cm]{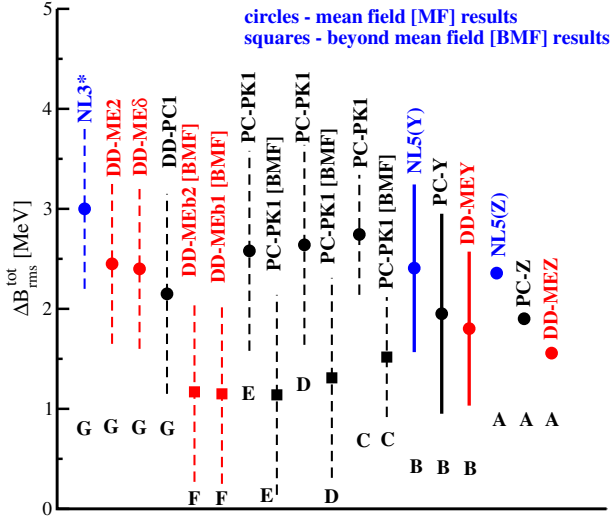}
\caption{The summary of existing global mass calculations which satisfy the 
condition $\Delta B_{rms} \leq 3.0$ MeV.  The labels A, B, C, D, E, F, G
are used for references \cite{NL5Z-DDMEZ-PCZ}, \cite{TA.23}, \cite{RHB-e-e-continuum.22}, 
\cite{YWZL.21}, \cite{LLLYM.15}, \cite{AGC.16} and \cite{AARR.14}, respectively. Red, 
blue and black colors are employed for the DDME, NLME and PC functionals, respectively.
See text for details.  
\label{fig-with-errors}
      }
\end{figure*}

  The importance of global calculation errors is illustrated in Fig.\ \ref{fig-with-errors}.
Prior to this paper and Ref.\ \cite{NL5Z-DDMEZ-PCZ}  the global calculation errors were neglected 
and the quality of the functionals were judged solely by comparing the circles and squares 
representing $\Delta B_{rms}$ in Fig.\ \ref{fig-with-errors}. In the Z class of the functionals 
the $\delta B_{rms}^{negl}$ values are purely numerical in  nature and are evaluated at the 
level of 28,  23 and 50 keV for the DD-MEZ, NL5(Z) and PC-Z  functionals, respectively. Note 
that respective errors bars are smaller than the size of the symbols in Fig.\ \ref{fig-with-errors}.  
One can clearly see that the best [worst] results are obtained with the DD-MEZ [NL5(Z)] 
functionals.

  The fitting protocols of the DD-MEZ and DD-MEY [as well as of NL5(Z) and
NL5(Y)] functionals are almost the same: a minor difference in the number of even-even 
nuclei used does not play a principal role.  Thus, the results obtained with the  Z functionals 
can be used as a benchmark to define the $\delta B_{rms}^{negl}$ values for the Y 
functionals. Note that the Y(Z) classes of the functionals are fitted in finite $N_F = 20$ 
and $N_B = 20$ (infinite) bases with neglect (accounting) of total electron binding 
energies. As a result, the optimum minima of the Z and Y types of the functionals
are found in different locations of the parameter hyperspace. Thus, the 
binding energies obtained with  DD-MEY [NL5(Y)] in a given nucleus deviate (by 
up to several MeV) from exact ones obtained with DD-MEZ [NL5(Z)]. This mechanism 
leads to  $\delta B_{rms}^{negl}=0.77$ MeV and 0.84 MeV for the DD-MEY and NL5(Y)
functionals, respectively [see Table \ref{table-global-mass}]. For the PC-Y functional 
the analysis leads to somewhat larger $\delta B_{rms}^{negl}$ value (see Ref.\ 
\cite{NL5Z-DDMEZ-PCZ} for details) which reflects very slow convergence of the binding
energies as a function of $N_F$ in the PC functionals (see Refs.\ 
\cite{TOAPT.24,NL5Z-DDMEZ-PCZ}).

   No accurate assessment 
of global calculation errors due to the use of finite fermionic and bosonic bases and neglect 
of TEBEs exists for other global mass calculations
presented in Table \ref{table-global-mass}.  However, considering the similarity of their 
fitting protocols  and global calculations to that of the Y class of the functionals,  it is 
reasonable to assume that $\delta B_{rms}^{negl}$ is of the order of 0.8 MeV for the CDFT 
mass tables based on the DDME and NLME functionals shown in Table  
\ref{table-global-mass}. 
The global calculation error is expected to be larger for the PC functionals. This
is due to three factors. First, it is difficult to obtain the binding energies corresponding 
to infinite fermionic basis in actinides and superheavy nuclei (see Refs.\ 
\cite{TOAPT.24,NL5Z-DDMEZ-PCZ}). Second, 
the contributions to the $\Delta B^{cor}(Z,N)$  quantity due to TEBEs and infinite basis 
corrections in the fermionic sector act in the same direction. Third,  the HO basis 
with $N_F'=12$, 14 and 16 fermionic shells  is used in the calculations of 
Refs.\ \cite{ZNLYM.14,LLLYM.15,YWZL.21} for the nuclei with $Z<20$, $20\leq Z <82$ 
and $82\leq Z \leq 104$ regions,  respectively. However, the PC-PK1 functional has been 
fitted  in truncated basis with $N_F=20$  in Ref.\ \cite{PC-PK1}.  The RHB calculations 
comparing the binding energies obtained with the $N_F'$ values in above discussed 
regions of nuclear chart with those employing $N_F=20$ clearly show that this difference 
in the basis size contributes additional 0.199 MeV to $\delta B_{rms}^{negl}$. Taking into 
account all these factors one can conclude that $\delta B_{rms}^{negl}[{\rm PC}]\approx 1.0$  
MeV is very likely the conservative (lowest) estimate for global calculation error in the case 
of the PC functionals.

    Fig.\ \ref{fig-with-errors} summarizes the situation and  shows that global calculation 
errors are quite substantial for most of the functionals and that they have to be taken into 
account when two functionals are compared. Moreover, they are unacceptable if one 
aims at the accuracy of the global description of binding energy at the level of
 $\Delta B_{rms}\approx 0.5$ MeV which is achieved in the state-of-the-art non-relativistic
models.  This clearly indicates the need for a proper accounting of total electron binding
energies and infinite basis corrections to binding energies in the fermionic sector of 
the CDFT. Note also that the analysis carried out in the Skyrme DFT shows that the neglect  of TEBEs
leads to substantial differences in binding energies as compared
with those obtained in accurate calculations which include this contribution  (see supplemental
material to Ref.\ \cite{PS.22}).
 
  One can also ask a question of what would be true value of  $\Delta B_{rms}$ 
if global calculation error is drastically reduced. In that respect it is interesting to compare 
the results obtained with the Woods-Saxon (WS) and  harmonic oscillator (HO) bases for the 
PC-PK1 functional 
presented in Refs.\  \cite{LLLYM.15,YWZL.21,RHB-e-e-continuum.22}. The calculations with the HO 
basis provide $\Delta B_{rms} = 1.14$ MeV \cite{LLLYM.15} and $\Delta B_{rms} = 1.31$  
 MeV \cite{YWZL.21} but the global calculation error $\delta B_{rms}^{negl}$  is estimated to 
be at least  1 MeV.
The calculations with the WS and HO bases neglect total electron binding energies and this 
generates substantial contribution to global calculation error. However, the truncation of 
the basis effects are substantially smaller in the WS basis of  Ref.\ \cite{RHB-e-e-continuum.22} 
than those  in the HO one.
Thus, one can  estimate that the  global calculation error in Ref.\ \cite{RHB-e-e-continuum.22} 
is  smaller (by approximately 40\%) than that in Refs.\ \cite{LLLYM.15,YWZL.21}. Despite that the 
calculations in the WS basis lead to a higher 
$\Delta B_{rms} = 1.518$ MeV (see Ref.\ \cite{RHB-e-e-continuum.22}). 
 It is interesting that  this value  obtained in the calculations 
with BMF effects taken into account is 
quite comparable with  $\Delta B_{rms}^{tot} = 1.557$ MeV 
 obtained for the DD-MEZ functional at the MF level.
 
\section{Unresolved  issues in global mass calculations}
\label{sec-common-issues}

\begin{figure*}[htb]
\centering
\includegraphics[width=13.0cm]{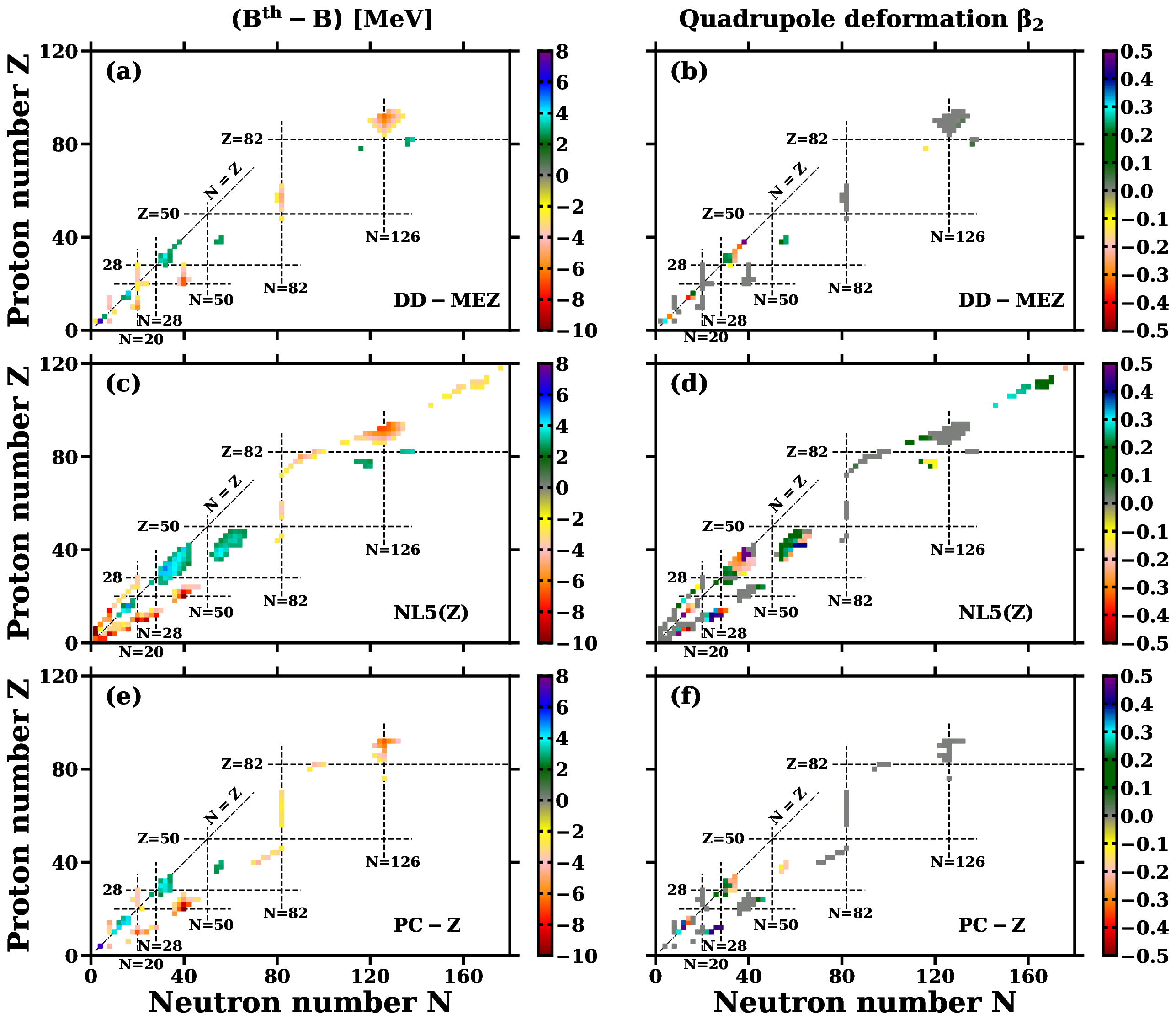}
  \caption{(left panels) The differences $B^{th} - B $ between calculated ($B^{th}$) 
and experimental $(B)$ nuclear binding  energies obtained with indicated functionals. 
Note that only the nuclei which satisfy the condition $|B^{th} - B| \geq 2.5 $ MeV are 
shown here.  (right panels) The calculated quadrupole deformations  $\beta_2$ of the 
nuclei shown on respective left panels. 
\label{common-issues}
      }
\end{figure*}
 
   It is important to understand whether unresolved common issues exist in the global 
description of binding energies with three classes of CEDFs under study.
Fig.\ \ref{common-issues} displays the nuclei for which the difference between
theory and experiment exceeds 2.5 MeV. It clearly reveals the existence of similar 
problems in the description of binding energies with different classes of CEDFs.

    First, let us consider the nuclei which are shown in green or blue color in the 
left column of Fig.\ \ref{common-issues}. Such nuclei are less bound in the calculations than 
in  experiment. One group of such nuclei is located in the vicinity of the $N=Z$ line. For them
the  neglect of the Wigner energy is a possible reason for substantial discrepancies between
theory and experiment. It provides an extra binding for the $N=Z$ nuclei and  its inclusion could 
reduce the difference between theory and experiment. However, to our knowledge it is neglected 
in the fitting protocols of all existing CEDFs and in the majority of non-relativistic ones since its 
microscopic origin and explicit form are still under debate (see Refs.\ 
\cite{SDGMN.97,GSHPT.02,Neer.09,BF.13,FM.14} and  references quoted therein).  
The addition of the Wigner energy in the formulation of Ref.\ \cite{GSHPT.02} reduces 
$\Delta B_{rms}$ from 1.601 MeV down to 1.557 MeV in the 
case of the DD-MEZ functional (see Ref.\ \cite{NL5Z-DDMEZ-PCZ}): the final result is shown in Fig.\
\ref{fig-ddmez}. 

   Another group of the nuclei in which the calculations underestimate the binding energy most
likely represents the cases of shape coexistence in which the relative energies of spherical 
and deformed minima or oblate and prolate minima are not properly reproduced.  The most
pronounced is the case of  two regions centered around $N\approx Z \approx 36$ and
$Z\approx 42, N\approx 60$ in the calculations with NL5(Z), see Fig.\ \ref{common-issues}(c).
Theoretical uncertainties in the predictions of the position in particle numbers at which the 
transition from one type of shape to another takes place are non-negligible (see discussion 
of Fig. 18 in Ref.\ \cite{AARR.14} and Fig. 7 in Ref.\ \cite{TAA.20}). Moreover, the calculations
at the mean field level not always correctly reproduce the shape of the nucleus in the ground
state in the case when competing minima coexist (see detailed discussion of charge radii in 
the Zr, Sr, Kr and Hg  isotopic chains in Ref.\ \cite{PAR.21} and references quoted therein). 
The accounting of beyond mean field effects may improve the description of shape coexistence 
in these regions and  provide needed additional binding via configuration mixing.

    The nuclei shown in red or pink color in Fig.\ \ref{common-issues} are more bound 
in the calculations than in experiment. There are three regions common to  considered classes  
of the CEDFs where such nuclei appear.  The first region is centered around $Z = 92, N=126$. In 
this region, the nuclei (which are spherical or near-spherical [see right column in Fig.\ 
\ref{common-issues}]) are substantially more bound in the calculations than in experiment and 
this over-binding  increases on getting closer to its center. They act as outliers which have 
disproportional impact on $\Delta B_{rms}$:  in the case of the DD-MEZ functional the $\Delta B_{rms}$
improves from 1.557 MeV  to 1.403 MeV if the nuclei in this region are removed from
consideration. The existence of this region is most likely due to the fact
the CEDFs based on the Hartree approach 
overpredict the size of the $Z=92$ spherical shell gap (see Refs.\ 
\cite{LSGM.07,PCF-PK1}). 
The second (less pronounced) region is located 
along the $N=82$ line for the $Z$ values near 58. Moderate overbinding 
seen in the calculations in this region as compared with experiment may be
due to the fact that employed functionals generate proton $Z=58$ spherical 
shell gap which is larger than the $Z=64$ one in contrast to experiment which 
indicates the presence of only $Z=64$ gap (see Refs.\ \cite{LSGM.07,PCF-PK1}).

Note that the comparison of experimental and calculated spectra for above discussed 
gaps  is done in Refs.\  \cite{LSGM.07,PCF-PK1} at the mean field level, i.e. particle-vibration 
coupling (PVC) which can significantly affect single-particle properties is ignored.
However,  the PVC calculations of $^{208}$Pb based on the NLME NL3* 
functional clearly indicate the presence of the $Z=92$ shell gap located
between the $1h_{9/2}$ and $2f_{7/2}$ spherical subshells which is larger
by a factor of $\approx 2$ than experimental one (see Fig. 5 in Ref.\ \cite{LA.11}). 
Note that the size of this gap is substantially reduced in the PVC calculations as 
compared with the one obtained at the mean field level.

   Finally, the third region 
is located along the neutron drip-line or its vicinity: it consist of two subregions  
located around  $Z\approx 10, N\approx 28$ and $Z\approx 22, N\approx 40$. 
In this region, substantial discrepancies between calculated and experimental
binding energies are at least partly caused by the deficiencies in 
the description of the size of the $Z=8, 20$ and $N=28, 40$ spherical shell
gaps (see discussion below).

  The right column of Fig.\ \ref{common-issues}  presents calculated quadrupole
deformations $\beta_2$. Let start their analysis from the best performing DD-MEZ and 
PC-Z functionals. One can see that for these CEDFs the absolute majority of 
the  nuclei shown on figure are spherical in the ground
states. This suggests that the discrepancy between calculations and experiment 
is caused by the deficiencies in the description of the energies of spherical 
subshells and the sizes of spherical shell gaps.
For these two functionals a number of nuclei shown in Figs.\ \ref{common-issues}(b)
and (f) also possess quadrupole deformation but they are mostly located at or near 
the $N=Z$ line.   As discussed above, the neglect of the Wigner energy is a possible 
reason for their presence in this figure. These features are also seen for the NL5(Z)
functional but for a somewhat larger number of nuclei due to its worse performance. 
In addition, quite substantial deviations between theory and experiment exist
for this functional in two regions centered around $N\approx Z \approx 36$ and 
$Z\approx 42, N\approx 60$ (see Fig.\ \ref{common-issues}(c)) in which prolate-oblate 
shape coexistence exists (see Fig.\ \ref{common-issues}(d)) and which  is quite likely
not  properly described (see discussion above).

Note that indicated above issues are common for the CEDFs developed at the Hartree 
level. This was verified by a global comparison between theory (at the mean field level) and 
experiment not only for the above mentioned functionals but also for the DD-MEX, DD-MEY 
(see Fig. 1 in Ref.\ \cite{TA.23}),  
DD-ME2,  DD-PC1, DD-ME$\delta$, NL3* (see Fig.\ 6 in Ref.\ \cite{AARR.14}), and PC-PK1 (see Fig. 
3(a) in Ref.\ \cite{LLLYM.15} and Fig. 1 in Ref.\ \cite{YWZL.21}).  The inclusion of the 
correlations beyond mean field (for example, in the framework of five-dimensional collective 
Hamiltonian) improves the accuracy of the global description of masses but does not 
eliminate above mentioned local problems (see Fig.\ 3(b) in Ref.\ \cite{LLLYM.15} and Fig. 
1(b) in Ref.\ \cite{YWZL.21}).

\section{Conclusions}

   The status of global calculations of binding energies within the covariant density 
functional theory has been reviewed. For many years, they were almost exclusively
based  on the functionals which were fitted to restricted set of spherical nuclei. However,
such fitting protocols create  a global bias.  It is only with advent of anchor based 
optimization approach in Ref.\ \cite{TA.23} that global optimizations of
CEDFs in a large fermionic basis or in the basis accurately approximating infinite 
one become possible.
 
  In model calculations the contributions leading to smooth deviations  trends between 
approximate and accurate solutions as a function of particle number are frequently ignored 
under assumption that they can be accounted  via the redefinition of the parameters of the 
functional. Such approximations are utilized in fitting protocols of relativistic 
and non-relativistic EDFs and they are a reflection of the limitations of our knowledge and 
theoretical tools. The current study reveals substantial deficiencies of such an approach since it 
leads to global  calculation errors in binding energies.
For example, such errors are of the order  of 1 MeV when total 
electron binding energies and infinite basis corrections to binding energies in fermionic sector 
(for the $N_F=20$ basis) are neglected.
 
   The global optimizations of CEDFs have been carried out by our group at the mean field level. 
The implementation of the BMF effects at phenomenological level via rotational and vibrational 
corrections is in progress and it is expected to further improve the global accuracy of the 
description of binding energies. However, such corrections are not expected 
to eliminate the problems in the description of the nuclei which are more bound in the calculations 
as compared with experiment at the mean field level (the nuclei which are shown in red or
pink colors in the left column of Fig.\ \ref{common-issues}).   Thus, the present analysis indicates the 
need for further  improvement of spectroscopic quality of the CEDFs at the Hartree level via global 
investigations of the impact  of other interaction channels such as 
tensor interaction \cite{MEK.23,SP.24}  and $\delta$-meson \cite{DD-MEdelta}
 on the single-particle energies. 
\\
\\
This material is based upon work supported by the U.S. Department of Energy,  Office of Science, 
Office of Nuclear Physics under Award No. DE-SC0013037.

\bibliography{references-46-optimization-HO-basis.bib}

\end{document}